# Hybrid Optoelectronic Correlator Architecture for Shift Invariant Target Recognition

Mehjabin Sultana Monjur,[1] Shih Tseng,[1] Renu Tripathi,[3] John James Donoghue,[4] and M.S. Shahriar[1,2,*]

[1]Department of Electrical Engineering and Computer Science, Northwestern University, Evanston, IL 60208, USA
[2]Department of Physics and Astronomy, Northwestern University, Evanston, IL 60208, USA
[3]OSCAR, Department of Physics & Engineering, Delaware State University, Dover, DE 19901, USA
[4]Digital Optics Technologies, Rolling Meadows, IL 60008
[*]Corresponding author: shahriar@northwestern.edu

In this paper, we present theoretical details and the underlying architecture of a hybrid optoelectronic correlator that correlates images using Spatial Light Modulators (SLM), detector arrays and Field Programmable Gate Array (FPGA). The proposed architecture bypasses the need for nonlinear materials such as photorefractive polymer films by using detectors instead, and the phase information is yet conserved by the interference of plane waves with the images. However, the output of such a Hybrid Opto-electronic Correlator (HOC) has four terms: two convolution signals and two cross-correlation signals. By implementing a phase stabilization and scanning circuit, the convolution terms can be eliminated, so that the behavior of an HOC becomes essentially identical to that of a conventional holographic correlator (CHC). To achieve the ultimate speed of such a correlator, we also propose an Integrated Graphic Processing Unit which would perform all the electrical processes in a parallel manner. The HOC architecture along with the phase stabilization technique would thus be as good as a CHC, capable of high speed image recognition in a translation invariant manner.

## 1. Introduction

Target identification and tracking is important in many defense and civilian applications. Optical correlators provide a simple technique for fast verification and identification of data. Over recent years, we have been investigating the feasibility of realizing an all-optical high-speed automatic target recognition (ATR) correlator system using the inherent parallelism of optical techniques[1-7]. Other groups have also pursued the development of such correlators [8,9]. The simplest form of such a system is the basic Vander Lugt [10] optical correlator which is illustrated schematically in fig. 1. Here, each lens has a focal length of L. The process starts with an image (e.g., the reference image, possibly retrieved from a holographic memory disc or with a Spatial Light Modulator (SLM) connected to a computer) in the input plane $P_1$. The reference image passes through the lens producing Fourier Transform of the image at plane $P_M$. Now a plane wave is applied, at an angle φ in the y-z plane, to interfere with the fourier transformed image in the plane $P_M$. The interference is recorded in a thin photographic plate, which produces a transmission function that is proportional to the interference pattern. Once the recording is fixed, the query image (e.g., from a camera connected to an SLM) is presented in the input plane $P_1$. After passing through the first lens, the fourier transform of the query image passes the photographic plate. After passing through the second lens, the cross-correlation signal is observed at the output plane, $P_2$. The amplitude of this cross-correlation signal is high when the reference image and the query image are matched. If the query image is shifted with respect to the input image in the x-y plane, the correlation spot will also appear shifted. The other signals produced (for example, the convolution between the two images) during this process do not overlap with the cross-correlation signal if φ is chosen to be sufficiently large. The limitations of such an architecture is that the recording process is very time consuming. This constraint is circumvented in a Joint Transform Correlator (JTC), where a dynamic material such as photorefractive polymer film is used so that the recording and correlation take place simultaneously. A JTC makes use of a dynamic non-linear material, such as photorefractive thin film produced by Nitto-Denko [1,2]. The primary limitation of this system is the poor nature of the material used for making the JTC. First, it is very fragile, and gets destroyed rather easily for reasons not well understood. Second, the diffraction efficiency is rather small, and it produces a lot of scattering, leading to a very poor signal-to-noise (SNR) ratio. Third, after carrying out some correlations, the residual gratings generated in the medium have to be erased by applying a high voltage; this process takes quite a bit of time. To our knowledge, all dynamic holographic films suffer from similar limitations.

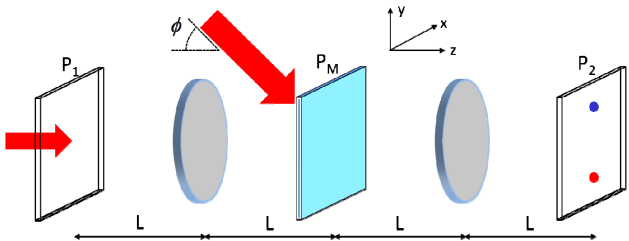

Fig 1: Schematic illustration of a conventional optical correlator

The scheme presented in this paper gets around this problem by making use of concepts developed recently in the context of digital holography[10]. Namely, the non-linearity provided by the JTC medium is replaced by the non-linearity of high speed detectors (since detectors measure intensities, they are inherently non-linear). Of course, this requires some modification of the architecture, as well as post-processing of signals. In this paper, we specify an explicit and novel architecture that enables the process of correlating images using detectors only, and the phase information is conserved by interfering with plane waves. We would like to point out that Javidi et al. [11] demonstrated a JTC that also makes use of detectors, in a manner similar to what we described here. However, this approach has some key limitations. The reference and the query images have to be placed in the same field of view. In order to do so, it is necessary to convert each image to a digital format first, and then create a composite image, which is sent to an SLM, for example. This process precludes the use of a scenario, necessary for very rapid search, where the reference images are retrieved directly from a holographic memory disc. For this scenario, it would be necessary to combine the reference image, which is in the optical domain, with the query image, also in the optical domain after generation from an SLM, using a beam splitter. When this is done, the correlation will depend sinusoidally on the relative phase difference between the two optical paths, with no obvious option for stabilizing or optimizing this phase difference. Our approach presented here circumvents both of these constraints.

The paper is organized as follows: Section 2 presents the theoretical details of the proposed Hybrid Optoelectronic Correlator (HOC). The phase stabilization and scanning technique, which eliminates the convolution terms, is also described in this section. Section 3 presents some possible future works on realizing an optoelectronic processor that would speed up the whole process of correlation. Section 4 describes simulation results of the proposed HOC using MATLAB, illustrating its performance for various cases of interest.

## 2. Proposed Hybrid Optoelectronic Correlator

The overall architecture of the HOC proposed here is summarized in fig. 2 and fig. 3. Briefly, the reference image, $H_1$ is retrieved from the database and transferred to an optical beam using an SLM (SLM-1), and is Fourier transformed using a lens. The FT image ($M_1$) is split into two identical ports. In one port, the image is detected by an array of detectors, which could be a high resolution

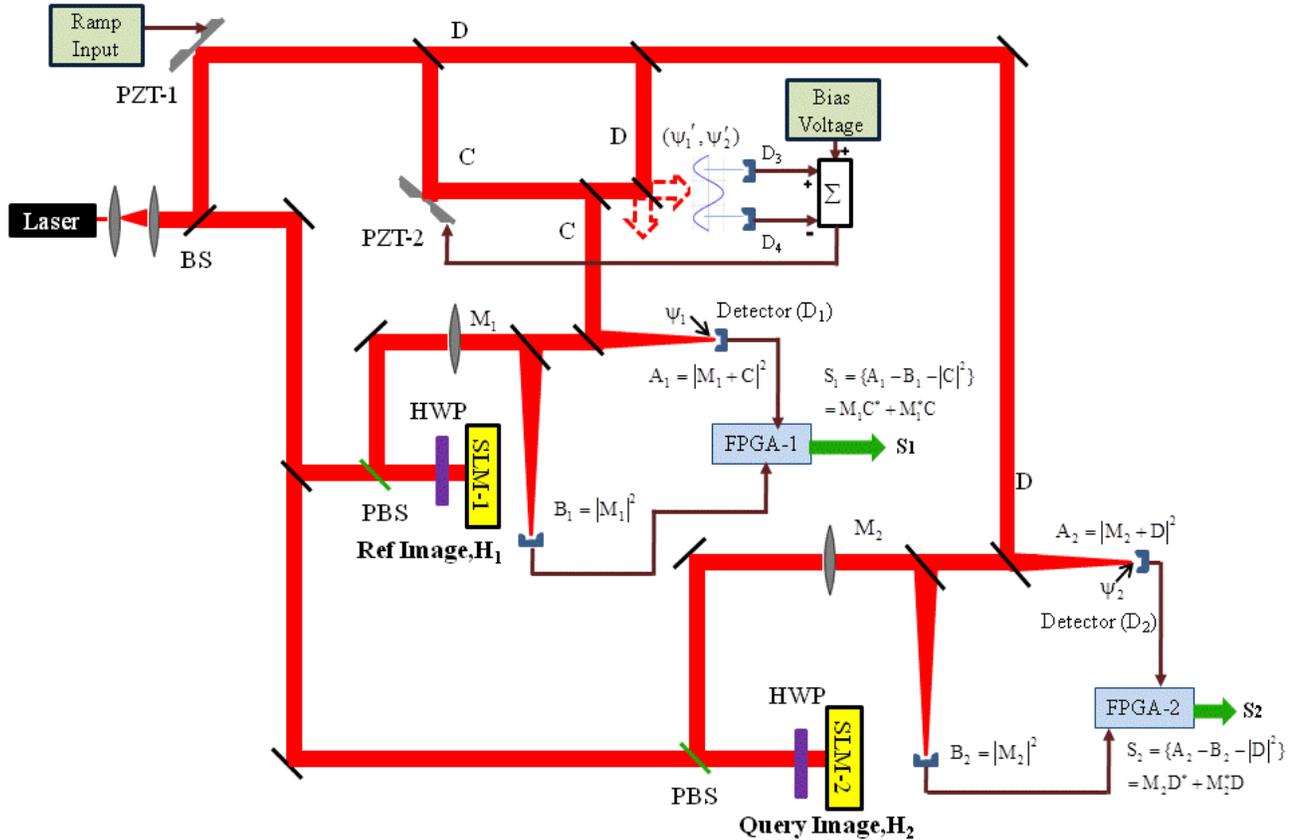

Fig. 2: Proposed architecture of the hybrid optoelectronic correlator with phase stabilization and scanning circuit; BS = Beam Splitter; PBS = Polarizing Beam Splitter; HWP = Half Wave Plate; The dotted lines pertain to discussions in section 3 about alternative, high speed implementation. (please keep this figure unchanged)

focal plane array (FPA) or a digital CMOS camera. As an explicit example, we consider the USB2.0 CMOS Camera (DCC1545M), which has 1280H x 1024V pixels and sends 10 bit data of each pixel at a 48 MHZ clock rate, thus requiring about 27 ms to send an image. The signal array produced by the camera is denoted as $B_1$. The camera is interfaced with a Field Programmable Gate Array (FPGA) via a USB cable. $B_1$ can be stored in the built in memory of the FPGA [FPGA-1]. In the other port, $M_1$ is interfered with a plane wave C, and detected with another CMOS Camera, producing the digital signal array $A_1$ and is stored in the memory of FPGA-1. $A_1$ & $B_1$ can be expressed as:

$$A_1 = \left| M_1 e^{j\phi_1} + C e^{j\psi_1} \right|^2$$
$$= |M_1|^2 + |C|^2 + |M_1||C|e^{j(\phi_1-\psi_1)} + |M_1||C|e^{-j(\phi_1-\psi_1)} \quad (1)$$

$$B_1 = |M_1|^2 \quad (2)$$

In addition, the intensity profile of the plane wave ($|C|^2$) is measured, by blocking the image path momentarily, using a shutter (not shown), and the information is stored in the memory component of FPGA-1. FPGA-1 then computes and stores $S_1$ which can be expressed as:

$$S_1 = A_1 - B_1 - |C|^2 = M_1 C^* + M_1^* C$$
$$= |M_1||C|e^{j(\phi_1-\psi_1)} + |M_1||C|e^{-j(\phi_1-\psi_1)} \quad (3)$$

Here $\phi_1(x,y)$ is the phase of the Fourier transformed image, $M_1$, and $\Psi_1$ is the phase of the plane wave, C. It should be noted that $\phi_1$ is a function of (x,y), assuming that the image is in the (x,y) plane. This subtraction process has to be done pixel by pixel using one or more subtractors available in the FPGA. Consider, for example, the Virtex-6 ML605 made by Xilinx as a candidate FPGA, which has an oscillator frequency of 200 MHz, so that each subtraction takes about 5-10 ns depending on the implementation of the adder circuit. Thus, the total subtraction process of an image of size 1280 x 1024 pixels would take about 6 ms when the subtraction is done with one subtractor. This process can be speeded up by using multiple subtractors that can operate in parallel in many FPGAs.

The captured query image, $H_2$ is transferred to an optical beam using another SLM (SLM-2), and split into two paths after being Fourier transformed with a lens. The resulting image in each path is designated as $M_2$. In a manner similar to what is described above for the query image, the signal $S_2=A_2-B_2-|D|^2$ is produced using two cameras and an FPGA (FPGA-2) and stored in FPGA-2 memory. Here, D is the amplitude of an interfering plane wave, and the other quantities are given as follows:

$$A_2 = |M_2 + D|^2$$
$$= |M_2|^2 + |D|^2 + |M_2||D|e^{j(\phi_2-\psi_2)} + |M_2||D|e^{-j(\phi_2-\psi_2)} \quad (4)$$

$$B_2 = |M_2|^2 \quad (5)$$
$$S_2 = M_2 D^* + M_2^* D = |M_2||D|e^{j(\phi_2-\psi_2)} + |M_2||D|e^{-j(\phi_2-\psi_2)} \quad (6)$$

As before, $\phi_2(x,y)$ is the phase of the Fourier transformed image, $M_2$, and $\Psi_2$ is the phase of the plan wave D.

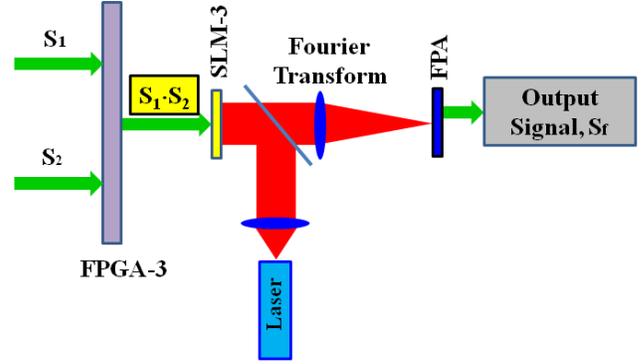

Fig 3: The final stage of the Hybrid Optoelectronic Correlator

In the final stage of the hybrid correlator (as shown in fig 3), these two signals ($S_1$ and $S_2$) described in equations (3) and (6) are multiplied together using the multiplier in FPGA-3. Four quadrant multiplication can easily be implemented using an FPGA. The resulting signal array, S is stored in FPGA-3 memory. This can be expressed as:

$$S = S_1 \cdot S_2 = \left(M_1 C^* + M_1^* C\right)\left(M_2 D^* + M_2^* D\right)$$
$$= \left[\alpha^* M_1 M_2 + \alpha M_1^* M_2^* + \beta^* M_1 M_2^* + \beta M_1^* M_2\right] \quad (7)$$

where, $\alpha \equiv |C||D|e^{j(\psi_1+\psi_2)}$; $\beta \equiv |C||D|e^{j(\psi_1-\psi_2)}$.

This signal array, S, is now transferred to another SLM (SLM-3) from FPGA-3 through the Digital Visual Interface (DVI) port. For the X-Y series SLM made by Boulder Nonlinear Systems (BNS), for example, this image update would take about 65 ms. Since S can be positive or negative, the SLM should be operated in a bipolar amplitude mode. The optical image produced by SLM-3 is Fourier transformed using a lens, and detected by an FPA. The output of the FPA will provide the main correlation signal. The final signal can thus be expressed as:

$$S_f = \left[\alpha^* F(M_1 M_2) + \alpha F(M_1^* M_2^*) + \beta^* F(M_1 M_2^*) + \beta F(M_1^* M_2)\right] \quad (8)$$

Here, F stands for the Fourier Transform and $\alpha, \beta$ are constants. Since $M_j$ is the FT of the real images $H_j$, j=1,2, using the well-known relations between the FT of products of functions, and convolutions and cross-correlations, we can express the final signal as the sum of four terms:

$$S_f = \alpha^* T_1 + \alpha T_2 + \beta^* T_3 + \beta T_4 \quad (9)$$

$T_1 = H_1(x,y) \otimes H_2(x,y)$
$T_2 = H_1(-x,-y) \otimes H_2(-x,-y)$
$T_3 = H_2(x,y) \odot H_1(x,y)$
$T_4 = H_1(x,y) \odot H_2(x,y)$

Where $\otimes$ indicates two-dimensional convolution, and $\odot$ indicates two-dimensional cross-correlation. We can now make the following observations:
- $T_1$ represents the two-dimensional convolution of the images, $H_1$ and $H_2$.
- $T_2$ represents the two-dimensional convolution of the images, $H_1$ and $H_2$, but with each conjugated and inverted along both axes.

- $T_3$ represents the two-dimensional cross-correlation of the images, $H_1$ and $H_2$.
- $T_4$ represents the two-dimensional cross-correlation of the images, $H_2$ and $H_1$. (Cross-correlation is non-commutative; hence, $T_3$ is not necessarily equal to $T_4$)
- If the images, $H_1$ and $H_2$, are not symmetric in both x and y directions, we have $T_1 \neq T_2 \neq T_3 \neq T_4$
- If the images, $H_1$ and $H_2$, are symmetric in both x and y directions, we have $T_1 = T_2 = T_3 = T_4$

The cross-correlation technique is usually used to find matches between two objects. In our final result we have convolution terms ($T_1$ & $T_2$) in addition to cross-correlation terms ($T_3$ & $T_4$). The convolution terms can be washed out by implementing a phase stabilization and scanning technique in the HOC architecture.

*Phase Stabilization & Scanning Circuit:*

From eqn 9, it is obvious that the final signal $S_f$ depends nontrivially on $\Psi_1$ & $\Psi_2$. To make this dependence more transparent, we can rewrite eqn 7 as follows:

$$S = 2|C||D||M_1||M_2| \left[ \cos(\psi_1 + \psi_2 - \phi_M) + \cos(\psi_1 - \psi_2 - \tilde{\phi}_M) \right] \quad (10)$$

where, $M_1 M_2 = |M_1||M_2| e^{j\phi_M(x,y)}$ and $M_1 M_2^* = |M_1||M_2| e^{j\tilde{\phi}_M(x,y)}$. Here the first term corresponds to the convolution and the second term corresponds to the cross-correlation. To eliminate the convolution term, we continuously scan ($\Psi_1 + \Psi_2$) over a range of $2\pi$ at a certain frequency $\omega_s$, while keeping ($\Psi_1 - \Psi_2$) zero. The convolution term varies as we scan ($\Psi_1 + \Psi_2$), whereas the cross-correlation term remains constant (since $\Psi_1 - \Psi_2 = 0$). While scanning is going on, we pass the signal S through a low pass filter (LPF) with a bandwidth less than $\omega_s$, so that the last term which corresponds to cross-correlation is passed. Such a filter can be easily implemented with the FPGA. The low passed filtered version of S is Fourier transformed using a lens and detected by an FPA, producing only the cross-correlation signals.

Fig. 2 shows the architecture for phase stabilization and scanning, where ($\Psi_1 - \Psi_2$) is kept to a constant value using a simple interferometer along with a feedback loop and ($\Psi_1 + \Psi_2$) is scanned over a range of $2\pi$ using a Piezoelectric Transducer (PZT). To start with, the output of the laser is split into two paths. One path is used for generating images via reflections from SLM (as already shown in fig 2). The other path is used for generating both reference beams, C & D. Using a PZT (PZT-1) mounted on a mirror before the beam is split into C & D, and applying a saw-tooth type ramp voltage on it, result in a repeated linear scan of ($\Psi_1 + \Psi_2$) over a range of $2\pi$. Pieces of C & D are now split off (not shown earlier in fig 2 for simplicity) and interfered with each other. The resulting interference pattern along with a feedback signal applied to another PZT (PZT-2) mounted on a mirror in the path of C, can be used to control the value of ($\Psi_1 - \Psi_2$). To be specific, note that, $\Psi_1 - \Psi_2 = \xi + (\Psi_1' - \Psi_2')$ where, $\Psi_1$ & $\Psi_2$ are the phases at the detectors $D_1$ and $D_2$, respectively, $\Psi_1'$ & $\Psi_2'$ are the phases of C and D at the detectors of the interferometer ($D_3$ & $D_4$) and $\xi$ is a constant, determined by path length difference. This condition of ($\Psi_1 - \Psi_2$) = 0 can be achieved if ($\Psi_1' - \Psi_2'$) = -$\xi$. We describe below how this can be achieved.

The interference signal is detected by a pair of matched detectors ($D_3$ & $D_4$). The voltages from these detectors are subtracted from each other using a subtractor circuit. The resultant voltage from the subtractor is added with a bias voltage using an adder circuit and the output of this adder is fed to PZT-2. The feedback loop can lock the interference pattern at any desired position. By performing several correlation operations of two identical images with this setup at different bias voltages, we find the bias voltage that gives the maximum peak value of the cross-correlation signal. At this position, ($\Psi_1' - \Psi_2'$) = -$\xi$, which makes ($\Psi_1 - \Psi_2$) = 0.

With the system locked at this position, ($\Psi_1 + \Psi_2$) is varied over a range of $2\pi$ by applying a ramp voltage to PZT-1, as mentioned above. The response of the feedback loop should be faster than $\omega_s$, in order to ensure that the servo can hold ($\Psi_1' - \Psi_2'$) to a constant value. While the scanning is going on, the signal is passed through an LPF with a bandwidth less than $\omega_s$. This low pass filtered version of S is then processed to yield cross-correlation signals only, as discussed in section 2.

While the image detection process is going on, the stability of the phases should be checked after some characteristic time, $T_c$. This characteristic time is defined as the time during which ($\Psi_1 - \Psi_2$) can drift within a certain allowable range, for example, a few milliradians. After time $T_c$, we have to adjust the bias voltage again and perform several correlation of two known images with the HOC to get the highest correlation peaks. The characteristic stability time would depend on the stability of the optical mounts, and can easily exceed 100's of seconds in a well designed system.

## 3. Possible Future Work to Speed up Operation of HOC

In describing the architecture of the HOC in section 2, we have considered the use of commercially available components such as cameras, FPGAs and SLMs. However, it is obvious from the analysis that the overall process is severely slowed down during the serial communication between these devices. In order for the HOC to achieve its ultimate operating speed, it is thus necessary to resort to novel components that operate in parallel.

Consider the set of steps (shown in fig. 2) whereby signals captured by the cameras are processed to produce the signal S, which then appears as an optical field at the output of the final SLM (SLM-3). This process is serving as a conduit between signals that start in the optical domain (i.e. inputs for the cameras) and end in the optical domain (i.e. output of the SLM-3). With proper use of current technologies, it should be possible to combine these tasks in an integrated graphic processing system for high-speed operation, as shown in fig 4(a). Fig. 4(b) shows the block diagram of the specialized integrated graphic processing unit (IGPU) that combines, in parallel, the FPA, the ASIC (Application-specific Integrated Circuit) chip for signal processing, and the high-speed SLM.

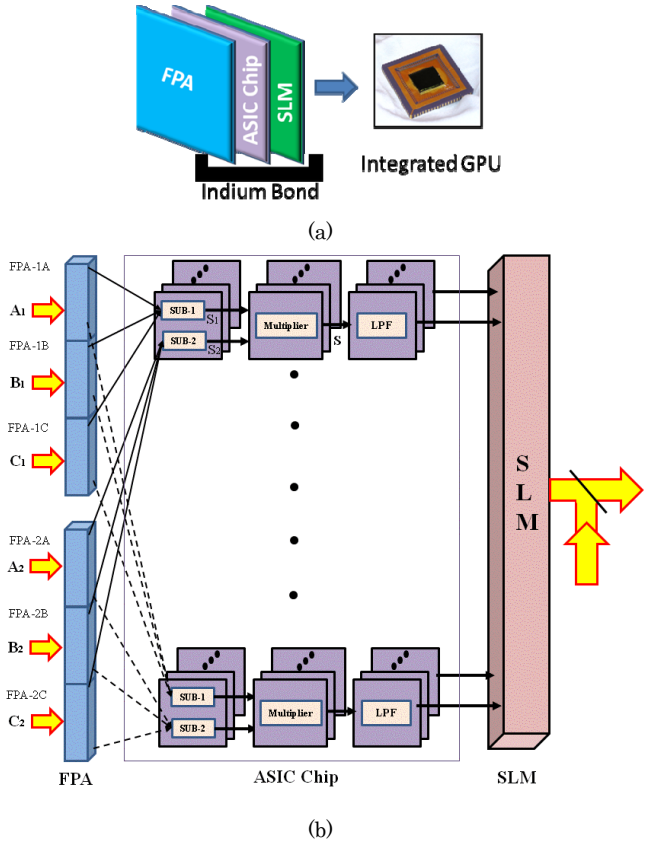

Fig 4: (a) Shows the block diagram of the integrated graphic processing unit (IGPU) (b) Shows the block diagram of the specialized IGPU in detail that combines, in parallel, the FPA, the ASIC (Application-specific Integrated Circuit) chip for signal processing, and the high-speed SLM. [SUB = Subtractor; FPA= Focal Plane Array; LPF = Low Pass Filter; GPU= Graphic processing Unit, SLM=Spatial Light Modulator ]

Briefly, FPA-1A, FPA-1B and FPA-1C would capture, respectively, signals, $A_1(\equiv|M_1+C|^2), B_1(\equiv|M_1|^2)$ and $|C|^2$, corresponding to the reference channel [12]. Similarly, FPA-2A, FPA-2B and FPA-2C would capture, respectively, signals, $A_2(\equiv|M_2+D|^2), B_2(\equiv|M_2|^2)$ and $|D|^2$, corresponding to the query channel. The FPA would be connected to the ASIC chip through Indium bonding, for example. The ASIC chip would consist of 512 X 512 signal-processing units, corresponding to each pixel and geometrically matched to the FPA. Each unit of the ASIC chip would have two analog subtractor circuits, one analog multiplier circuit and an analog low pass filter (LPF). The signals from the matching pixels in the three FPAs (FPA-1A, FPA-1B and FPA-1C) will be processed by a single subtractor element (denoted as SUB-1) in the ASIC chip, producing the signal $S_1= A_1 - B_1 - |C|^2$. Similarly, FPA-2A, FPA-2B and FPA-2C, along with SUB-2, would generate the signal array $S_2$, corresponding to the query channel. These two signals, $S_1$ and $S_2$, would be applied in parallel, to an analog multiplier, producing the signal array S. The signal S is passed through an LPF to get rid of the convolution terms, as discussed in section 2. On the back end of the ASIC chip array would be the SLM array, connected via indium bonding.

The speed of operation of such an IGPU would, of course, depend on the specific technology employed. In what follows, we describe a specific example of the technology that could be employed to realize each part of the IGPU, thus enabling us to reach a definive estimate of the speed of operation.

Consider first the FPA. One possible choice for this would be an array of nano-injection detectors [13,14]. These detector elements operate at a low bias voltage (~ 1 V) at room temperature, and has a response time of 1 ns. Consider next the subtractor elements. Each of these could be implemented easily with a simple operational amplifier consisting of as few as six transistors[15]. The response time of such an operational amplifier is expected to be similar (~0.3 us) to that of a bulk operational amplifier chip, such as LM741C. The multiplier could be realized with the Gilbert cell, which requires only six transistors [13]. The properties of such a multiplier should be similar to that of a bulk multiplier chip, such as AD835 which is a complete four-quadrant, voltage output analog multiplier. It generates the linear product of its X and Y voltage inputs with a rise time of 2.5 ns. The LPF can be implemented with 0.1 nf capacitance and 10 ohm resistance would have a response time of 1ns. Finally, the SLM could be realized with an array of high speed stepped quantum wells [16-20]. An SLM based on these elements has a response time of ~16 ns, which is orders of magnitude faster than the conventional SLM discussed in section 2.

Therefore, the integrated graphic processing unit would take less than 0.4 µs to perform the whole process of capturing optical signals, computing the signal S and converting it back to optical domain. In contrast, for performing the same operation of 512X512 image, the technology discussed in section 2 takes about 22 ms since the data communication is done serially. Thus, the IGPU would lead to a speed-up of operation by a factor of nearly $5*10^4$. Of course, we have simply outlined a sketch of the type of the integrated graphic processing unit that is required to make the HOC achieve its ultimate speed. We are working with collaborators to design and implement this chip, and results from these efforts will be reported in the future.

## 4. Results of Numerical Simulations

As discussed in section 2, the convolution terms can be washed out by scanning continuously ($\Psi_1 + \Psi_2$) over a range of $2\pi$ at a certain frequency $\omega_s$, while keeping ($\Psi_1 - \Psi_2$) zero. This can be verified through some simulation results shown in fig 5. Two identical but shifted images (as shown in fig. 5(a)) are inputs to the HOC architecture without phase stabilization circuits (the HOC architecture as shown in fig. 2). When phase stabilization and scanning technique is not implemented and $\Psi_1$ & $\Psi_2$ are both kept zero, FPA detects a signal $|S_f|^2$ which contains both convolution ($T_1$ & $T_2$) and cross-correlation ($T_3$ & $T_4$) terms, as shown in fig 5(b). Another case without phase stabilization and scanning technique is shown in fig 5(c)

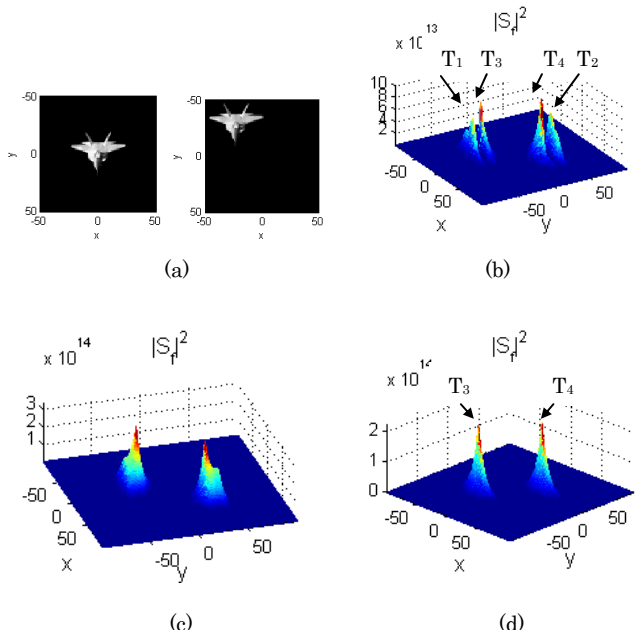

Fig 5: (a) Two identical but shifted images are inputs to the HOC. (b) All four terms are present in the output of the detector when $\Psi_1 =\Psi_2= 0$ (without phase stabilization circuit). (c) The output of the detector, $|S_f|^2$ when $\Psi_1 = \Psi_2 = \pi/3$ (This result is also without phase stabilization circuit) (d) Phase stabilization and scanning gives the optimum value of $|S_f|^2$ by eliminating the convolution terms ($T_1$ & $T_2$) and keeping only the cross-correlation terms ($T_3$ & $T_4$).

where $\Psi_1$ & $\Psi_2$ are both set to the value of $\pi/3$. In this case also, the convolution terms appear in $|S_f|^2$ along with the cross-correlation terms. Next we vary ($\Psi_1 + \Psi_2$) from 0 to $2\pi$, in 20 intervals, and average the values of S for these 20

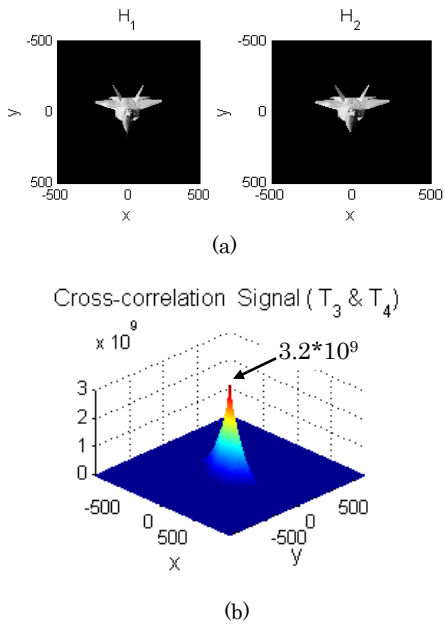

Fig. 6: (a) Two similar images with same position, $H_1$ & $H_2$ are the inputs to the HOC (b) The cross-correlation signals ($T_3$ & $T_4$) give sharper peak due to matched images

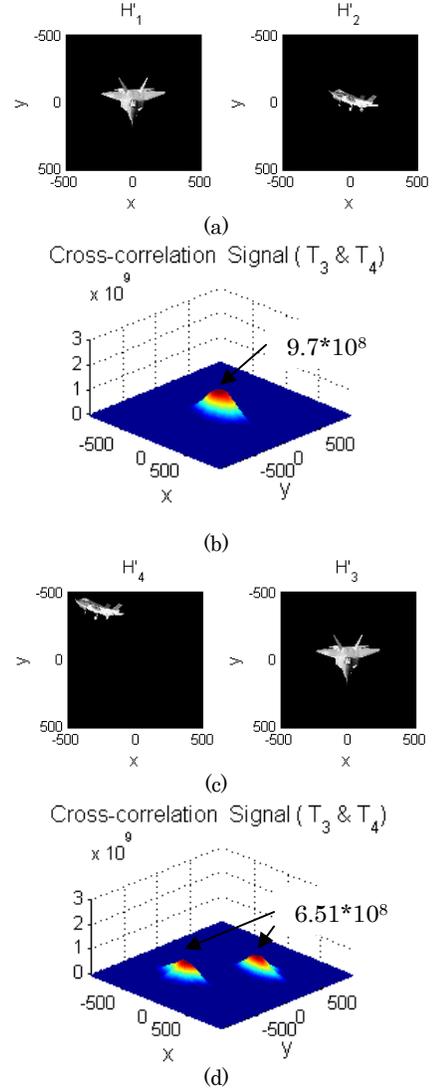

Fig. 7: (a) Two completely different images $H'_1$ and $H'_2$ are inputs to the HOC (b) The cross-correlation terms, $T_3$ & $T_4$ do not reveal any sharper peak, since they are different. (c) Two different and shifted images, $H'_3$ and $H'_4$ are inputs to the HOC. (d) The cross-correlation terms also ($T_3$ & $T_4$ ) do not reveal sharper peak.

cases. This is, of course, equivalent to the process of low pass filtering with a bandwidth less than the inverse of the duration for a linear scan of ($\Psi_1 + \Psi_2$) over a range of $2\pi$, as described in the section 2. As shown in fig 5(d), this averaging eliminates the convolution terms, and $|S_f|^2$ contains only the cross-correlation terms. In what follows, we will assume that such an averaging/filtering is carried out, and plot only the cross-correlation terms

Next, we illustrate the behavior of the HOCarchitecture for various types of objects and reference images. In fig 5(d) we have already shown the result for identical but shifted images. Fig 6(a) shows the case where the reference image, $H_1$ and the query image, $H_2$ are identical and not shifted. The cross correlation signals ($T_3$ & $T_4$ ) are shown in fig 6(b). In this case, the cross correlation signals ($T_3$ & $T_4$ ) have a peak at the center of the detector plane.

Fig. 7(a) shows another case, where two different images, $H_1$ and $H_2'$, are the inputs to the HOC. In this case, as seen in fig 7(b), the cross-correlation signals ($T_3$ & $T_4$) do not reveal any distinct peak. Fig 7(c) shows the case of two different images, $H_3'$ and $H_4'$, shifted with respect to each other. In this case, the cross-correlation terms (fig 7(d)) also do not have sharp peaks, and the maximum value is less than that shown in fig 7(b). Of course, it is also much smaller than the matched cases shown in fig 5(d) and 6(b).

It is important to note that if the query and reference images are not shifted with respect to one also another, then the cross-correlation peaks will be at the center of the detection plane, overlapping each other. If the images are shifted with respect to one another, the cross-correlations will be shifted symmetrically around the center by a distance corresponding to the shift.

## 5. Conclusion

We have presented theoretical details and the underlying architecture of the HOC that correlates images using SLMs, detector arrays and FPGAs. The HOC architecture bypasses the need for photorefractive polymer films by using detectors, and the phase information is yet conserved by the interference of plane waves with the images. The output signal of such a HOC has four terms: two convolution signals and two cross-correlation signals. The convolution terms can be eliminated by implementing a phase stabilization and scanning circuit, so that the behavior of an HOC becomes essentially identical to that of a CHC. To achieve the ultimate speed of such a correlator, we also propose an opto-electronic chip which would perform all the electrical processes in a parallel manner. The HOC architecture along with the phase stabilization technique would thus be as good as a CHC, capable of high speed image recognition in a shift invariant manner. In addition to shift invariant property of the HOC, rotation and scale invariant correlation can also be achieved by applying Polar Mellin Transform (PMT) to both query and reference images [21]. With the future implementation of an opto-electronic chip and PMT, the HOC architecture holds the promise of a practical, versatile and high speed image recognition system.

In this paper, we have not considered the issue of what the typical signal to noise raio (SNR) would be for the HOC architecture. Obviously, the SNR would depend on the details of the implementation. Experimental efforts are underway in our group to demonstrate the operation of the HOC, and the issue of the SNR as well other possible practical constraints would be addressed in the context of reporting on the outcome of this experimental effort.

## 6. Acknowledgement

This work is supported by AFOSR Grant FA9550-10-01-0228, NSF CREST award 1242067 and NASA URC Group-V award NNX09AU90A.